\documentclass[12pt]{article}
\usepackage[textures]{graphics}
\parskip=\bigskipamount
\newcounter{itemlistc}
\newcounter{romanlistc}
\newcounter{alphlistc}
\newcounter{arabiclistc}
\newenvironment{itemlist}
{\setcounter{itemlistc}{0}
\begin{list}{$\bullet$}
{\usecounter{itemlistc}
\setlength{\parsep}{0pt}
\setlength{\itemsep}{0pt}}}{\end{list}}

\def\qed{\hbox{${\vcenter{\vbox{ 
   \hrule height 0.4pt\hbox{\vrule width 0.4pt height 6pt
   \kern5pt\vrule width 0.4pt}\hrule height 0.4pt}}}$}}

\newcommand{\beq}{\begin{equation}}
\newcommand{\eeq}{\end{equation}}
\newcommand{\beqa}{\begin{eqnarray}}
\newcommand{\eeqa}{\end{eqnarray}}
\newcommand{\ket} [1] {\vert #1 \rangle}
\newcommand{\bra} [1] {\langle #1 \vert}
\newcommand{\braket}[2]{\langle #1 | #2 \rangle}

\begin{document}

\centerline{\bf A new solution for the Mean King's problem.}

\vspace*{0.37truein}
\centerline{\footnotesize Thomas Durt}
\vspace*{0.015truein}
\centerline{\footnotesize\it TENA-TONA Theoretical
 Physics Vrije Universiteit Brussel Pleinlaan 2} 
\baselineskip=10pt
\centerline{\footnotesize\it 1050 Brussels Belgium.}

\vspace*{0.21truein}
{\it Abstract: When the state of a quantum system belongs to a $N$-dimensional
 Hilbert space, with $N$ the power of a prime number, 
it is possible to associate to the system a finite field (Galois field) with $N$ elements.  
In this paper, we introduce generalized Bell states that can be intrinsically expressed 
in terms of the field operations.
   These Bell states are in one to one correspondence with
    the $N^2$ elements of the generalised Pauli group or Heisenberg-Weyl group. 
    This group consists of discrete displacement operators and 
    provides a discrete realisation of the Weyl function.
     Thanks to the properties of generalised Bell states and of quadratic extensions of finite
      fields, 
     we derive
    a particular solution for the Mean King's problem. 
    This solution is in turn shown to be in one to one correspondence with a set of $N^2$  self-adjoint
     operators that provides
a discrete realisation of the Wigner quasi-distribution.}



\section{ Introduction} 
\noindent
We showed previously,\cite{DurtNagler} in the framework of quantum cloning, that in dimension
 $N$ = 4 different classes of Bell states can be defined, that are associated to different groups of permutations 
of the $N$ basis states. These Bell states were shown to be in one to one correspondance
 with a commutative group
 (generalised addition) of $N$ elements. More recently,\cite{Durtmutu}
  we showed thatin dimension 4 there exists
  a multiplication that together with the addition forms a finite field or Galois field
   and that the generalised Bell states can be defined intrinsically in terms of the field properties.
   
 We also showed that whenever the dimension of the Hilbert space is a
  prime power ($N=p^m$ with $p$ a prime and $m$ a positive integer), these properties can be generalised.  
  This is due to the fact that finite fields (sometimes called division rings)
   with
   $N$ elements only exist when $N$ is a prime power.

  Besides, it is known\cite{Wootters,india} that when the dimension is a prime power there exists a maximal set of $N+1$ mutually unbiased\cite{Schwinger,Ivanovic}
   bases or MUB's
    (two orthonormal bases of a $N$ dimensional Hilbert space are said to be mutually 
  unbiased if whenever we choose one state in the first basis, and a second state in the second basis,
   the modulus squared of their in-product is equal to $1/N$).
  We showed in Ref.\cite{Durtsept} how to (re)derive in a straightforward
   manner an expression for the states of these bases, in agreement with the expressions of 
   Wootters and Fields\cite{Wootters}
    and with the approach of Bandyopadhiay\cite{india} et al. .
    This expression can be intrinsically formulated  in terms of the field operations
     (addition and multiplication between
     $N$ elements) and of the additive characters of the field ($p$th root of unity when $N=p^m$). 
  
   The existence of $N+1$ MUB's is directly related to the so-called
    Mean King's problem\cite{vaid,Englert,2003} which can be formulated as follows, in the qubit case ($N=2$):
    
    {\bf A Mean King challenges a physicist, Alice, who got stranded on the remote island ruled by the King,
  to prepare a spin1/2 atom in any state of her choosing and to perform a control measurement of her liKing.
   Between her preparation and her measurement, the King's men determine the value of either 
 $\sigma_{X}$, $\sigma_{Y}$ or $\sigma_{Z}$. Only after she completed the control measurement,
  the physicist is told which spin component has been measured, and she must then state the result of that intermediate
   measurement correctly. How does she do it?}
   
   In other words, how is it possible to ascertain the spin values of the spin observables along the $X$, $Y$ and $Z$ directions?
   The corresponding eigenbases are MUB's, so that the problems consists of ascertaining the values of $2+1$
    non degenerate observables of which the eigenbases are mutually unbiased relatively to each other.
    
   In prime power dimension ($N=p^m$), the Mean King's problems consists
    of finding a way to
     ascertain the values of $N+1$ non-commuting and non-degenerate observables that
     are diagonal in the $N+1$ mutually unbiased bases.
   
   We shall show in the present paper a way to generalise to prime power dimensions the solutions that can be
    found in the litterature for the Mean King's problem in dimension 2 and in prime
     dimensions, making use of the properties of the generalised Pauli group, of the 
   generalised Bell states, and of their transformation law. This solution is shown
    to be a special case of the general solution obtained by Aravind\cite{2003}. 
    
    An appealing property of our particular solution is that it provides a discrete 
   counterpart of Wigner's distribution that is valid in even and odd prime power dimensions as well. 
   We shall also show in appendix how it is possible to generalise 
   certain of these results in arbitrary odd dimensions, provided we reformulate the Mean King's problem in a
    slightly different manner.

\section{Preliminary concepts}

\noindent
Whenever the dimension $N$ of a Hilbert space 
  is a prime power ($N=p^m$), with $p$ a prime number, and $m$ a positive integer,
   it is possible to associate to the Hilbert space a finite field with $N$ elements\footnote{A field is a set with a multiplication and an addition operation which satisfy the usual rules, associativity and commutativity
    of both operations, the distributive law, existence of an additive identity 0 and a multiplicative identity 1, additive inverses, and multiplicative 
    inverses for every element, 0 excepted. }.
   
   We shall in the following label the elements of the field by a $m$uple of integer numbers
     $(i_{0},i_{1},...,i_{m-1})$ 
  running from 0 to $p-1$. This $m$uple $(i_{0},i_{1},...,i_{m-1})$ is in turn in one to one correspondence with
   an integer 
  number $i$, $0\leq i\leq N-1$) that we define by its $p$-ary expansion as follows
  
   $i=\sum_{k=0}^{m-1}i_{n}p^n$.
   
    In what follows we shall make no difference between elements of the fields and their 
    integer counterpart, to the contrary of usual conventions.
     
  Any finite field is characterized by two operations, a multiplication and an addition, that we shall denote 
 $\odot_G$ and $\oplus_G$ respectively. 

   The index $G$ refers to Evariste Galois and is introduced in order not to confuse
   these operations with the 
 usual (complex) multiplication and addition 
  for which no index is written.

 It is not a simple problem to find the multiplication table of finite fields, 
 but mathematicians solved the problems and such tables are available ``on line''.

 Besides, we can assign an integer label to the elements of the field in such a way that when the
  elements of the field are in one-to-one correspondence with $m$uples (as explained before)¥,
  the addition is equivalent with the addition modulo $p$ componentwise. 
  Actually this procedure corresponds to what is called a choice of basis for the field. It imposes certain contraints on the 
  identification between integers and elements of the field.
  For instance, we must assume in order to avoid inconsistencies that 0 corresponds 
  to the neutral element for addition. It is a natural choice to associate 1 to the neutral element 
     for multiplication; then , the $m-1$ remaining ¥powers of $p$¥
      are assigned to elements of the field
      that are independent in the same way that elements of a vectorial space are independent. ¥¥

  As a consequence of fact that the addition is equivalent to the addition modulo $p$ componentwise¥, one can show that whenever we add an element of the field $p$times with itself we obtain 0.
   In particular, the characteristics of the field,
   which is the smallest number of times that we must add the element 1 
   (neutral for the multiplication) with itself before we obtain 0 (neutral for the addition),
    is always equal to a prime number
    ($p$ when $N=p^m$). This last property is valid for any finite field and ultimately explains why
     the number of elements of a
     finite field necessarily equals a power of prime.

Let us denote $\gamma_{G}$ the $p$th root of unity: $\gamma_{G} =e^{ i .2\pi/p}$. 
Exponentiating  $\gamma_{G}$ with elements $g$ of 
the field (with the usual rules for exponientation), we obtain complex phasors of the type 
$\gamma_{G}^{ g}$ ($0\leq g\leq N-1$). Such phasors can take $p$ different values. They can be
 considered as a $p$-valued 
generalisation of the (binary) parity operation $e^{ i .(2\pi/2).g}$ that 
corresponds to the qubit case in the sense
 that the phasor $\gamma_{G}^{ g}$ ($0\leq g\leq N-1$) only depends on the value of 
   the first component $g_{0}$ of the $p$-ary expansion of $g$. $g_{0}$ is nothing else
    than the remainder of $g$ after division by $p$,
    when the division by $p$ is taken in the usual sense. 
    
    Exponentiating $\gamma_{G}$, 
     the $p$th root of unity, by this number 
     we obtain a complex phase which is also equal to $\gamma_{G}$ exponentiated by the rest of 
     this number after division by $p$.
    
     It is worth noting that all the finite fields are equivalent, up to a relabelling,
  and that our choice of labelling (and of basis) is dictated by requirements of simplicity and convenience.
   Other approaches are also valid, for instance the ones that imply the field theoretical
    Trace.\cite{Wootters,india,klimov} Each approach presents its own advantages and disadvantages regarding
     practical applications. With our conventions, many computations are formally similar
      in prime and prime power dimensions, and certain expressions are simpler than in other approaches.
       The price to pay is a loss of generality, in the sense that our conventions can be shown to be equivalent to 
       the ones made in the Trace approach provided we perform a special
        choice of basis in the latter approach.\cite{Durtsept}

Making use of the fact that the addition is the addition modulo $p$, componentwise,
 we can derive the following identity which will appear to be very useful in the following:
 
 \begin{equation} \label{identi2}\gamma_G^{i}\cdot\gamma_G^{j}=
 \gamma _{G} ^{(i\oplus_G j)}\end{equation} 
 Indeed, $\gamma_G^{i}\cdot\gamma_G^{j}=\gamma _{G} ^{(i+ j)}=\gamma _{G} ^{(i_{0}+ j_{0})}=
 \gamma _{G} ^{(i\oplus_G j)_{0}}=\gamma _{G} ^{(i\oplus_G j)}$. This relation 
 expresses that $p$th roots of unity are additive 
    characters of the Galois field.\cite{Karpilovsky}
   
The following identity also appears to play a fundamental role in our approach:   
   \beq\label{identi1}\sum_{j=0}^{N-1} \gamma_{G}^{ (j\odot_{G} i)}=N\delta_{i,0}\end{equation} 

The proof is given in Ref.\cite{Durtsept}.
  
  It is important to note, in order to avoid confusions, that different types of operations are present at this level: the internal field operations are labelled
   by the lower index $G$. 
  They must not be confused with the modulo $N$ operations. It is worth noting that the property $\sum_{p=0}^{N-1} \gamma^{ (p\odot q)}=N\delta_{q,0}$ is true for the modulo $N$ multiplication as well,
    but $ \gamma$ must be taken to be equal to the $N$th root of unity in this case.
  In prime dimensions $\gamma _{G}$ is the $N$th root of unity and the Galois and modulo $N$ operations coincide. 
 In prime power but non-prime dimensions, this is no longer true.
  Because of this, certain applications are also valid even when the basic operations, addition and multiplication, 
  do not form a field. 
 This is the case for instance with the transformation law of the Bell states that can be derived
  in all odd dimensions, provided we operate with the usual (modulo) operations.
   In appendix 1 we show that this property opens the way to a reformulation of the Mean King's problem
    that is valid when the dimension is an odd number, not necessarily equal to the power of an odd prime.

\section{MUB's in prime power dimensions.}

\noindent
In Ref.\cite{Durtsept}, we obtain explicitly the following expression for MUB's in prime power
   dimensions:
   \beqa 
  \ket{  e_{k}^i}={ 1\over \sqrt N}\sum_{q=0}^{  N-1}\gamma_{G}^
  { \ominus_{G} q\odot_{G} k}
  (\gamma_{G}^{(
  (i-1)\odot_{G} q\odot_{G} q)})^{{1 \over 2}} \ket{  e_{q}^0}
  \label{synthetic}\eeqa
  where $\ket{  e_{k}^i}$ represents the $k$th basis state
   of the $i$th MUB ($k$ runs from 0 to $N-1$ and $i$ from 1 to $N$) expressed 
   in terms of the states of the 
  computational basis $ \ket{  e_{q}^0}$. The $N$ bases so-defined and the computational basis are mutually unbiased relatively to each other. 
  
  At this level there remains some ambiguity because a square root is always defined up to a global minus sign.
  We showed in Ref.\cite{Durtsept}
   how to rise this ambiguity and obtained the following determinations for the square root  factor $(\gamma_{G}^{(
  (i-1)\odot_{G} q\odot_{G} q)})^{{1 \over 2}}$:
  \begin{itemlist}
  \item in odd prime power dimensions, 
  $ (\gamma_{G}^{ (i-1)\odot_{G} q\odot_{G} q)})^{{1 \over 2}}=
  \gamma_{G}^{
  (i-1)\odot_{G} q\odot_{G} q)/_{G}2}$ (where $/_{G}$ represents division in the Galois
   field). The corresponding expression for the MUB's is thus, in odd prime power dimensions, 
   \begin{equation} 
  \ket{  e_{k}^i}={ 1\over \sqrt N}\sum_{q=0}^{  N-1}\gamma_{G}^
  { \ominus_{G} q\odot_{G} k}
  (\gamma_{G}^{(
  (i-1)\odot_{G} q\odot_{G} q)/_{G}2}) \ket{  e_{q}^0}
  \label{xxx},\end{equation} 
  
  \item in even prime power dimensions,
  
    $(\gamma_{G}^{\ominus( (j-1)\odot_{G} q\odot_{G} q)})^{{1 \over 2}}$=$
   (\gamma_{G}^{\oplus( (j-1)\odot_{G} q\odot_{G} q)})^{{1 \over 2}}$
   = $\Pi^{m-1}_{n=0, l_{n}\not= 0}i^{(j-1)\odot_{G}2^n\odot_{G}2^n}
  \gamma_{G}^{(j-1)\odot_{G}2^n\odot_{G}2^{n'}}$
  
  where $i$ represents $e^{{2\pi\over 4}}$, and where the coefficients $l_{n}$ are unambiguously defined by the $p$-ary (here binary) expansion of $l$: $l=\sum_{k=0}^{m-1}l_{n}2^n$, while 
 $n' $ is the smallest integer strictly larger than $n$ such that $l_{n'}\not= 0$, if it exists, 0 otherwise; here $\gamma_{G}=-1$¥.
 The corresponding expression for the MUB's is thus, in even prime power dimensions ($2^m$),
 \beqa
  \ket{  e_{k}^i}={ 1\over \sqrt N}\sum_{q=0}^{  N-1}\gamma_{G}^
  { \ominus_{G} q\odot_{G} k}
 \Pi^{m-1}_{n=0, q_{n}\not= 0}i^{(j-1)\odot_{G}2^n\odot_{G}2^n}
  \gamma_{G}^{(j-1)\odot_{G}2^n\odot_{G}2^{n'}}\ket{  e_{q}^0}
  \label{yyy},\eeqa
where $q=\sum_{k=0}^{m-1}q_{n}2^n$, while 
 $n' $ is the smallest integer strictly larger than $n$ such that $q_{n'}\not= 0$, if it exists, 0 otherwise.

  \end{itemlist}

   Although we proved in Ref.\cite{Durtsept} by direct computation that these bases are mutually
   unbiased, in the present paper we shall rather follow the approach of Ref.\cite{india} which is more general. 
  
   In the reference \cite{india}
     it is shown that when there exists a maximal commuting basis of orthogonal 
     unitary matrices, 
     the $N+1$ bases that diagonalize these classes are unambiguously defined and,
      moreover, are mutually unbiased. A maximal commuting basis of
      orthogonal unitary matrices is a set of $N+1$ sets of $N-1$ commuting unitary
       operators (or classes) 
      plus the identity such that these $N^2$ operators are orthogonal regarding the in-product
       induced by the (usual operator) trace denoted $tr.$. 
      Let us introduce the generalised displacement operators defined as follows:
      \beq
 V^j_i= \sum_{k=0}^{N-1} \gamma_{G}^{(( k\oplus_{G} i)\odot_{G} j)}\ket{ k\oplus_{G}
  i}\bra{  k}\label{defV0};i,j:0...N-1 \eeq
  It can be shown\cite{india,Durtsept} that the $N^2$ operators defined in this way constitute the so-called generalised Pauli group.

        They are unitary with $(V^j_i)^+=(V^j_i)^{-1}=
        \gamma^{\ominus_{G}(i\odot_{G} j)}
       V^{\ominus_G i}_{\ominus_G j}$, and are traceless operators (if we except the identity): $tr.V^j_i=N.\delta_{i,0}.\delta_{j,0}$. Besides, they obey the following composition law:
        \beqa
 V^j_i.V^k_l= \gamma^{\ominus_{G}(i\odot_{G} k)} V^{j\oplus_G k}_{i\oplus_{G} l}
 \label{compo} \eeqa 
 
  Making 
       use of the composition law  (\ref{compo}), it is straightforward to check that 
       the $N$ classes of operators $V_{l}^{(i-1)\odot_{G} l}$ 
       (where $l$ varies from 0 to $N-1$ and $i$ from 1 to $N$)
        contain $N-1$ commuting operators plus
        the identity, and that this is also true for the class $V_{0}^{l}$ (where $l$ varies from 0 to $N-1$). Besides, 
        the relation $tr.((V^j_i)^+.V^k_l)
       =N.\delta_{i,l}.\delta_{j,k}$ is valid so that the $N^2$ $V$ operators form a 
       maximal commuting basis of unitary operators. 
       According to the results established in the reference\cite{india} it is sufficient to find the 
       common eigenstates of these $N+1$ classes of $V$ operators 
       in order to determine the value of the 
       states of the $i$th MUB. Before we do so, it is useful to introduce
        the $U$'s operators: these operators are equal to the $V$'s operators,
         up to a well-chosen phase,
       \begin{equation} \label{conven}(\gamma_{G}^{\ominus(
  (i-1)\odot_{G} l\odot_{G} l)/_{G}2}) =U^{i}_{l}/V^{(i-1)\odot_{G}l }_{l}=
  (\gamma_{G}^{\ominus(
  (i-1)\odot_{G} l\odot_{G} l)})^{1\over 2}\end{equation} 
  We also impose that $U^0_{l}=V^l_{0}$.
 
 This phase is chosen in such a way that, inside a class of commuting operators the $U$'s operators obey an exact groupal 
 composition law,
 
 \beqa
U^{i}_{1_{1}}.U^{i}_{1_{2}} =  U^i_{l_{1}\oplus_G l_{2}}
 \label{compoU} \eeqa
 
 Actually it is the requirement of the fulfillment of an exact groupal composition law between
  the $U$'s operators that partially rises
  the ambiguity in the determination of the sign of the square root factor $(\gamma_{G}^{(
  (i-1)\odot_{G} q\odot_{G} q)})^{{1 \over 2}}$ (in prime odd dimensions
   our definition coincides with Weyl's definition\cite{weyl}). It was shown in Ref.\cite{Durtsept}
   that there exist exactly $N$¥
   different choices of
   the phases  $U^{i}_{1}/V^{(i-1)\odot_{G}l }_{l}$ that are consistent with
    the requirement that the composition law inside a subgroup of the generalised Pauli group 
   is exact. We also showed that there are $N$ different phase choices that
   ¥preserve the composition law and lead to bases that differ only by a Galois translation
    in the label of the basis states. 
    These choices are necessarily in one to one correspondence\footnote{Profs. Klimov and de Guise
    drew my attention on the problem caused by this ambiguity in relation with the different phase choices for
     the generalised Pauli group (¥¥during the ICSSUR conference hold
 in Besancon in May 2005).}. In a sense this is not astonishing because 
 the concept of MUB is independent on the ordering of the bases.

    Moreover, as a consequence of the composition law (\ref{compoU}), the following identity is satisfied:
  \beq(\gamma_{G}^{(j-1)\odot_{G}l_{1}\odot_{G}l_{1}})^{{1\over 2}}
 .(\gamma_{G}^{(j-1)\odot_{G}l_{2}\odot_{G}l_{2}})^{{1\over 2}}
 .\gamma^{(j-1)\odot_{G}(l_{1}\odot_{G} l_{2})}=(\gamma_{G}^{(j-1)\odot_{G}(l_{1}\oplus_{G}l_{2})
 \odot_{G}(l_{1}\oplus_{G}l_{2})})^{{1\over 2}}.\label{simili}\eeq
 Formally we can rewrite the previous equation as follows:
 $ (\gamma_{G}^{(j-1)\odot_{G}(a\oplus_{G}b)\odot_{G}(a\oplus_{G}b)})^{{1\over 2}} =
(\gamma_{G}^{(j-1)\odot_{G}(a\odot_{G}a)})^{{1\over 2}}.(\gamma_{G}^
{(j-1)\odot_{G}(b\odot_{G}b)})^{{1\over 2}}.(\gamma_{G}^{2.((j-1)\odot_{G}a
\odot_{G}b)})^{{1\over 2}}, $
 which is reminiscent of the equation (\ref{identi2}), although we are dealing here with 
half integer powers of $\gamma_{G}$ instead of integer powers.

Making use of the identity (\ref{simili}), we can now check 
       by direct substitution of the expression (\ref{synthetic}) that the states 
       $\ket{  e_{k}^i}$ are common eigenstates of the $i$th class ($i: 1...N$¥)¥: 
       \beqa
  V_{l}^{(i-1)\odot_{G} l}\ket{  e_{k}^i}=
  \sum_{k'=0}^{N-1} \gamma_{G}^{(( k'\oplus_{G} l)\odot_{G} (i-1)\odot_{G} l)}\ket{ k'\oplus_{G}
  l}\bra{  k}{ 1\over \sqrt N}\sum_{q=0}^{  N-1}\gamma_{G}^
  { \ominus_{G} q\odot_{G} k}
  (\gamma_{G}^{(
  (i-1)\odot_{G} q\odot_{G} q)})^{{1\over 2}} \ket{  e_{q}^0} \nonumber\\
  =
  { 1\over \sqrt N}\sum_{q=0}^{N-1} \gamma_{G}^{(( q\oplus_{G} l)\odot_{G} (i-1)\odot_{G} l)}
  \gamma_{G}^
  { \ominus_{G} q\odot_{G} k}
  (\gamma_{G}^{((i-1)\odot_{G} q\odot_{G} q)})^{{1\over 2}} \ket{  e_{q\oplus_{G}l}^0}\nonumber\\
  =
  \gamma_{G}^{  (l\odot_{G} k)}
   (\gamma_{G}^{( (i-1)\odot_{G} l\odot_{G} l))})^{{1\over 2}}{ 1\over \sqrt N}\sum_{q\oplus_{G}l=0}^{N-1}
  \gamma_{G}^{ \ominus_{G} (q\oplus_{G}l)\odot_{G} k}
  (\gamma_{G}^{((i-1)\odot_{G} (q\oplus_{G}l)\odot_{G} (q\oplus_{G}l))})^{{1\over 2}}
   \ket{  e_{q\oplus_{G}l}^0}\label{zzz}\eeqa
  The states of the computational basis are eigenstates of the operators of the 0th class.

\section{Bell states and MUB's.}

\noindent
There is a one-to-one correspondence between (generalized) Bell states  and
 the generalised Pauli group (\cite{DurtNagler,Durtmutu} see also \cite{Planat} 
 for a different approach based on additive and multiplicative characters of the Galois field).
  This correspondence is a key concept for explaining
 several important applications of quantum information science, such as quantum teleportation,
  quantum dense coding, quantum cloning, and 
 it also leads, combined to the properties of invariance of the Bell states
  in MUB's, to a solution of the Mean King's problem valid in prime power dimensions
   as we shall show in a next section.
  
Following  Refs.\cite{DurtKwek}, \cite{DurtNagler}, and \cite{Durtmutu},
 we can define the generalized Bell states as follows:
 
 \beqa
\ket{B_{m^*,n}}=N^{-1/2} \sum_{k=0}^{N-1} \gamma_{G}^{(k\odot_{G} n)}
\ket{k^*}\ket{k\oplus_{G}m}\label{bell}
\eeqa

In this definition, we introduced the basis states $\ket{k^*}$ which belong to the complex conjugate basis
 of the direct basis $\ket{k}$. This does not make any difference when $\ket{k}$ is 
 the reference (computational) basis but it does when the Bell states are defined relatively to a 
 basis that possesses states with complex amplitudes when they are expanded in the computational basis. Formally, 
 states of the form $\ket{k^*}$ transform like bra's when states $\ket{k}$ transform like kets. 
 
 According to the Eq.~(\ref{defV0}), $V^n_m=  \gamma_{G}^{(m\odot_{G} n)}.$$
 \sum_{k=0}^{N-1} \gamma_{G}^{(k\odot_{G} n)}\ket{ k\oplus_{G} m}\bra{  k}$.
 Up to a global phase and a normalisation factor, Bell states and displacement ($V$) operators are thus seen to be one and the same object. Bell states possess many interesting properties,
  they are maximally entangled and form an orthonormal basis of the 
 $N^2$ dimensional Hilbert space obtained by taking the tensor product of the $N$ dimensional Hilbert space with itself (system + ancilla)¥. 
 
 These Bell states possess two appealing symmetries in relation with the generalised Pauli group:
 
 \begin{itemize}\item the Bell states are invariant under the elements of the group (the $V$'s operators) (up to a global phase):
 \begin{equation} V^i_{j}\ket{ B^0_{m^*,n}}(V^i_{j})^{-1}= \gamma_{G}^{(m\odot_{G} i\ominus_{G}n\odot_{G} j)}.
 \ket{ B^0_{m^*,n}}\label{V} \end{equation} where the upper index refers to the reference basis relatively to which the Bell states are defined:
 
 \beqa \ket{ B^k_{m^*,n}}=N^{-1/2} \sum_{l=0}^{N-1} \gamma_{G}^{(l\odot_{G} n)}
\ket{  e^{k*}_{l}}\ket{  e^{k}_{l\oplus_{G}m}} \label{G}\eeqa

 The equality (¥¥\ref{V}) shows that (up to irrelevant, global phases) the 
 MUB's are eigen bases of
  a finite symmetry group: a set of transformations that preserves the Bell states (expressed in the computational basis)¥. 
 
 \item the Bell states are permuted among each other when they are reexpressed in any of the 
 MUB's:

\beqa \ket{ B^k_{m^*,n}} =\gamma_{G}^{(\ominus_{G}m\odot_{G} n)}
.(\gamma_{G}^{  ((k-1)\odot_{G}  n\odot_{G} n )})^{1\over 2}\ket{B^0_{ n^*,\ominus_{G}m\oplus_{G}((k-1)\odot_{G}n)}}, (k-1)=0...N-1\label{transfo}\eeqa
 \end{itemize}
 The proof is straightforward and is valid in even and odd prime power dimensions as well (in even prime power dimensions it is made use of the identity (\ref{simili}))¥:
 \beqa \ket{ B^k_{m^*,n}} =N^{-1/2} \sum_{l=0}^{N-1} 
 \gamma_{G}^{(l\odot_{G} n)}
\ket{  e^{k*}_{l}}\ket{  e^{k}_{l\oplus_{G}m}}\nonumber \\
 =N^{-1/2} \sum_{l=0}^{N-1}\gamma_{G}^{(l\odot_{G} n)}{ 1\over \sqrt N}\sum_{q=0}^{  N-1}\gamma_{G}^
  {  q\odot_{G} l}
  (\gamma_{G}^{(
  \ominus_{G}(k-1)\odot_{G} q\odot_{G} q)})^{{1 \over 2}} \ket{  e_{q}^0}
  { 1\over \sqrt N}\sum_{q'=0}^{  N-1}\gamma_{G}^
  { \ominus_{G} q'\odot_{G} (l\oplus_{G}m)}
  (\gamma_{G}^{(
  (k-1)\odot_{G} q'\odot_{G} q')})^{{1 \over 2}} \ket{  e_{q'}^0}\nonumber \\
  =N^{-1/2}\sum_{l,q,q'=0}^{N-1}{ 1\over  N}\gamma_{G}^
  { (\ominus_{G} q'\odot_{G} m)}\gamma_{G}^{(l\odot_{G} n)}\gamma_{G}^
  {  (q\ominus_{G} q')\odot_{G} l} (\gamma_{G}^{(k-1)\odot_{G} 
  (q'\odot_{G} q'\ominus_{G} q\odot_{G} q)})^{{1 \over 2}}
  \ket{  e_{q}^0}\ket{  e_{q'}^0}\nonumber \\
  =N^{-1/2}\sum_{l,q,q'=q\oplus_{G}i:0}^{N-1}{ 1\over  N}\gamma_{G}^
  { (\ominus_{G} (q\oplus_{G}i)\odot_{G} m)}\gamma_{G}^
  {  (n\ominus_{G} i)\odot_{G} l} (\gamma_{G}^{(k-1)\odot_{G} 
  ((q\oplus_{G}i)\odot_{G} (q\oplus_{G}i)\ominus_{G}\odot_{G} q\odot_{G} q)})^{{1 \over 2}}
  \ket{  e_{q}^0}\ket{  e_{q\oplus_{G}i}^0}\nonumber \\
  =N^{-1/2}\sum_{l,q,q'=q\oplus_{G}i:0}^{N-1}{ 1\over  N}N.\delta_{n\ominus_{G} i,0}
 \gamma_{G}^
  { (\ominus_{G} (q\oplus_{G}i)\odot_{G} m)}\gamma_{G}^{(k-1)\odot_{G} 
  q\odot_{G} i} (\gamma_{G}^{(k-1)\odot_{G} i\odot_{G} i})^{{1 \over 2}}
  \ket{  e_{q}^0}\ket{  e_{q\oplus_{G}i}^0}\nonumber \\
  = \gamma_{G}^{(\ominus_{G}m\odot_{G} n)}.(\gamma_{G}^{  ((k-1)\odot_{G}  n\odot_{G} n )})^{{1\over 2}}
\ket{B^0_{( n^*,\ominus_{G}m\oplus_{G}((k-1)\odot_{G}n)}}, (k-1)
=0...N-1\label{transfo} \qed\,.\eeqa

 This is a one-to-one mapping between the Bell states (up to global phases).
 
 Remark that as the Bell states are in one-to-one correspondence with the $V$'s operators, the corresponding transformation rule is valid for those operators: 
    $V^n_{m}(0)=phase.V^{m}_{\ominus_{G}n\oplus_{G}(i-1)\odot_{G}m}(i)$, where
   $V^n_{m}(0)=
  \sum_{k=0}^{N-1} \gamma_{G}^{(( k\oplus_{G} m)\odot_{G} n)}
  \ket{ e_{k\oplus_{G} m}^0}\bra{  e_{k}^0}$ and $V^n_{m}(i)=
  \sum_{k=0}^{N-1} \gamma_{G}^{(( k\oplus_{G} m)\odot_{G} n)}\ket{ e_{k\oplus_{G} m}^i}
  \bra{  e_{k}^i}; i:1...N$. In prime dimensions, the invariance of the generalised Pauli group under
   conjugation by any unitary matrix that maps the computational basis onto a 
   MUB is
    a basic property of a larger group that is known as the Clifford group and 
    has many applications in number theory and quantum computing.\cite{Appleby,davidsThesis}

 Now that we derived the transformation law of the Bell states, we have at our disposal nearly all the tools necessary
  in order to derive the solution of the Mean King's problem in prime power dimensions.

\section{ Solutions of the Mean King's problem.}

\noindent
\subsection{ The qubit case.}
\noindent
We shall firstly give an overview of the treatment in the simplest case (qubits).\cite{vaid} At first sight, the problem seems to be impossible to solve, because there exists no common eigenstate of 
the non-commuting observables $\sigma_{x}$, $\sigma_{y}$ and $\sigma_{z}$, and it is indeed impossible to discriminate between their 6 eigenstates. The solution consists in introducing an ancilla and of making use of the resource provided by entanglement.
The strategy of Alice is now to add an ancilla to the spinor that the King will measure and to prepare initially the maximally entangled state $\ket{B^{Z}_{0,0}}_{A,K}$.
 In virtue of the identity (\ref{transfo}), this state is invariant: $\ket{B^{Z}_{0,0}}_{A,K}$=$\ket{B^{X}_{0,0}}_{A,K}$=$\ket{B^{Y}_{0*,0}}_{A,K}$.
 When the King measures in the $k$th basis he projects the Bell state $\ket{B_{0*,0}}_{A,K}$ onto 
 $\ket{  e_{i}^{k*}}_{Alice}\otimes \ket{  e_{i}^k}_{King}, i=0...
 N-1=1,k=0...N=2 .$
  
Therefore Alice must, in order to save her head, be able to distinguish between the 6 product states
 $\ket{  e_{i}^{k*}}_{King}\otimes \ket{  e_{i}^k}_{Alice}, k=0...N=2,i=0,N-1=1 .$
  According to our previous conventions, the indices $k=$ 0,1 and 2 correspond to the $Z$, $X$ and $Y$ bases respectively.
  We rewrote in terms of the Bell states the solution derived by B-G Englert and Y. Aharonov\cite{Englert} which consists of the following: after the King performed his measurement, 
  Alice performs a von Neumann measurement in order to measure a non-degenerate observable which is diagonal in the 4-dimensional basis that is defined as follows:
  
  \beqa \label{MKbasis}\ket{\Psi}^{Z}_{1}={1\over 4}(\ket{B^{Z}_{0,0}}_{A,K}+\ket{B^{Z}_{0,1}}_{A,K}+\ket{B^{Z}_{1,0}}_{A,K}
  +i\ket{B^{Z}_{1,1}}_{A,K})\\ \nonumber
  \ket{\Psi}^{Z}_{2}={1\over 4}(\ket{B^{Z}_{0,0}}_{A,K}+\ket{B^{Z}_{0,1}}_{A,K}-\ket{B^{Z}_{1,0}}_{A,K}
  -i\ket{B^{Z}_{1,1}}_{A,K})\\ \nonumber
  \ket{\Psi}^{Z}_{3}={1\over 4}(\ket{B^{Z}_{0,0}}_{A,K}-\ket{B^{Z}_{0,1}}_{A,K}+\ket{B^{Z}_{1,0}}_{A,K}
  -i\ket{B^{Z}_{1,1}}_{A,K})\\ \nonumber
  \ket{\Psi}^{Z}_{4}={1\over 4}(\ket{B^{Z}_{0,0}}_{A,K}-\ket{B^{Z}_{0,1}}_{A,K}-\ket{B^{Z}_{1,0}}_{A,K}
  +i\ket{B^{Z}_{1,1}}_{A,K})\\ \nonumber\eeqa

According to our previous definitions, the four qubit Bell states are defined as follows:

\beq\label{Bell}\ket{B^{Z}_{m,n}}_{A,K}={1\over\sqrt{2}}\sum_{k=0}^{1}
(-)^{k.n}\ket{k}^Z_{A}\ket{k+_{mod 2}m}^Z_{K}\end{equation} where $m,n \in \{0,1\}.$
Consequently: \beqa
&&\ket{B^Z_{0,0}}_{A,K}={1\over \sqrt
2}\{\ket{0}^Z_{A}\ket{0}^Z_{K}+\ket{1}^Z_{A}\ket{1}^Z_{K}\} ,\qquad
\ket{B^Z_{0,1}}_{A,K}={1\over \sqrt
2}\{\ket{0}^Z_{A}\ket{0}^Z_{K}-\ket{1}^Z_{A}\ket{1}^Z_{K}\}\nonumber \\
&&\ket{B^Z_{1,0}}_{A,K} ={1\over \sqrt
2}\{\ket{0}^Z_{A}\ket{1}^Z_{K}+\ket{1}^Z_{1}\ket{0}^Z_{2}\},\qquad
\ket{B^Z_{1,1}}_{A,K}={1\over \sqrt
2}\{\ket{0}^Z_{A}\ket{1}^Z_{K}-\ket{1}^Z_{A}\ket{0}^Z_{K}\}
\nonumber\eeqa It is easy to check that the two last $\Psi^{Z}$ states are orthogonal to the product state
 $\ket{  e_{0}^{0}}_{Alice}\otimes \ket{  e_{0}^0}_{King}$,
 and the two first ones to $\ket{  e_{1}^0}_{Alice}\otimes \ket{  e_{1}^0}_{King}$.
 For instance $\braket{\Psi^{Z}_{3}}{  e^0_{0A}  e^0_{0K}}$ = $\braket{\Psi^{Z}_{1}}{  e^0_{1A}  e^0_{1K}}$
  = 0. Therefore,
  if Alice observes one of the two last (first) states and that afterwards the King reveals
   that he observed a state in the $Z$ basis she can infer unambiguously that the
    Mean King observed-measured-prepared the state 
   $\ket{  e_{1}^0}_{A}\otimes \ket{  e_{1}^0}_{K}$ ($\ket{  e_{0}^0}_{A}\otimes \ket{  e_{0}^0}_{K}$).
  Now, in virtue of the identity (\ref{transfo}), Bell states transform into Bell states when we pass from one of the three MUB's to another one:
  \beqa
\ket{B^{Z}_{0,0}}_{A,K}=\ket{B^{X}_{0,0}}_{A,K}=\ket{B^{Y}_{0^*,0}}_{A,K}\nonumber\\
\ket{B^{Z}_{0,1}}_{A,K}=\ket{B^{X}_{1,0}}_{A,K}=\ket{B^{Y}_{1^*,0}}_{A,K}\nonumber\\
\ket{B^{Z}_{1,0}}_{A,K}=\ket{B^{X}_{0,1}}_{A,K}=i\ket{B^{Y}_{1^*,1}}_{A,K}\nonumber\\
\ket{B^{Z}_{1,1}}_{A,K}=-\ket{B^{X}_{1,1}}_{A,K}=(-i)\ket{B^{Y}_{0^*,1}}_{A,K}\label{bla}\eeqa

So that the four states $\ket{\Psi}^{Z}$ are bijectively transformed in the four states $\ket{\Psi}^{X}$ and
 $\ket{\Psi}^{Y}$(up to unobservable phase changes!). Therefore
 Alice can infer without error the values of the spins along three orthogonal directions (a rather counterintuitive result!) and consequently
  save her head. It is worth noting that the solution expressed in Eq.~(\ref{MKbasis}) is equivalent to the one given in
   Refs.\cite{vaid,Englert} excepted that now it is formulated in terms of Bell states.

\subsection{The Mean King's problem in prime power dimensions.}
\noindent
We shall now consider the most general case (prime power dimensions) and show how
 the properties of invariance of the Bell states
   in MUB's also lead to a compact 
  expression of the solution, valid in all prime power dimensions.\cite{DurtMean}
 
  As before, Alice prepares initially the Bell state $\ket{B^{0}_{0,0}}_{A,K}$.
    The problem to solve can be formulated in this way: 
  in a Hilbert space of dimension $N^2$, with $N$ a prime power, is it possible to perform a single
   von Neumann measurement that allows us to discriminate between the product states
    $\ket{  e_{k}^{i*}}\ket{  e_{k}^{i}}$ ($k=0...N-1, i=0...N$)???. If this problem
     admits a solution, then it is possible to ascertain  the value of the spin of a spin $s$ particle
      (with $2s+1=N$) in $N+1$ MUB's.
      
      As in the qubit case, entanglement plays a central role in this approach, as well as the fact that there exists a finite field with $N$ elements when $N$ is a prime power. 
  We shall establish that a von Neumann measurement that satisfies all the constraints can be realised in the following basis: 
    \beqa \ket{\Psi}^{0}_{(i_{1},i_{2})}={1\over N}(\sum_{m,n=0}^{N-1}
  \gamma_{G}^{(i_{1},i_{2})\odot\odot_{G}(m,n)} (\gamma_{G}^{( m\odot_{G} n)})^{1\over 2} \ket{B^{0}_{m,n}}_{A,K})
  \label{alicestates}\eeqa where $\odot \odot_{G}$ represents the multiplication of the elements of the field with $N^2$ 
  elements. 
  This field is a quadratic extension of the field with $N$ elements and its elements can be represented by couples of elements of the field with $N$ elements. 
  The procedure of quadratic extension required to pass from a field with $N$ elements to a field with $N^2$ elements is
   similar by many aspects to the procedure of extension of the field of real numbers that leads to the complex field.\cite{Rubin}
It is easy to check the orthonormality of this basis, in virtue of the identity (\ref{identi1}):
$\sum_{m,n=0}^{N-1} \gamma^{ ((j_{1},j_{2})\odot\odot_{G}(m,n))}=N^2
\delta_{j_{1},0}\delta_{j_{2},0}$. 

Now, if the King prepares a product of states of the computational basis
 $\ket{  e_{l}^{0*}}_{Alice}\otimes \ket{  e_{l}^0}_{King}, l=0...N-1=1$, we have that

 $\braket{\Psi^{0}_{(i_{1},i_{2})}}{  e_{lA}^{0*}  e^0_{lK}}={1\over N}
 \sum_{m,n=0}^{N-1} \delta_{m,0}\gamma^{(\ominus_{G} i_{1}\odot_{G} m)}
 \gamma^{(\ominus_{G} i_{2}\odot_{G} n\odot_{G}
  R)}\braket{B^{0}_{0,n}}{  e^{0*}_{lA}
   e^0_{lK}}$  where $R$ is the remainder of $(0,1)\odot\odot_{G}(0,1)$ 
   after division by $N$. It can be shown that this remainder always differs from 0, as a consequence of the fact that a quadratic extension of a field is a field. 
   Finally, making use of $\braket{B^{0}_{0,n}}{  e^0_{lA}
   e^0_{lK}}={1\over \sqrt N}\gamma^{(\ominus_{G} n\odot_{G} l)}$ and of the identity (\ref{identi1}), we get that
     $\braket{\Psi^{0}_{(i_{1},i_{2})}}{  e^0_{lA}  e^0_{lK}}=
   {1\over \sqrt N}\delta_{\ominus_{G} i_{2}\odot_{G} R,l}$. This shows that,
   when a detector corresponding to any state that belongs to the basis 
 $\ket{\Psi}^{0}_{(i_{1},i_{2})}$ fires, we can infer unambiguously the value $l$ that was observed (prepared) by the King.
  In order to be able to infer the value of the King's
  observation/preparation in any MUB, we must have a similar relation when we reexpress the King's states and the  $\ket{\Psi}^{0}_{(i_{1},i_{2})}$
   state in the $k$th basis.  The transformation law for the $\ket{\Psi}$ states is the following: 
\beqa \ket{\Psi}^{k}_{(i_{1},i_{2})}={1\over N^2}(\sum_{m,n=0}^{N-1}\gamma_{G}^{(i_{1},i_{2})
\odot\odot_{G}(m,n)}
 (\gamma_{G}^{(\ominus_{G} m\odot_{G} n)})^{1\over 2} 
 .(\gamma_{G}^{  ((k-1)\odot_{G}  m\odot_{G} m )})^{1\over 2}
\ket{B^k_{(\ominus_{G}n\oplus_{G}(k-1)\odot_{G}m)^*,m}})
\label{transfowigner}\eeqa
 When the King prepares a product of states of the $k$th basis
 $\ket{  e_{l}^{k*}}_{Alice}\otimes \ket{  e_{l}^k}_{King}$, the preparation contains only superpositions of 
 Bell states  of the type $\ket{B^k_{0^*,m}}$, which imposes that $\ominus_{G}n\oplus_{G}(k-1)\odot_{G}m=0$. 
Thus $(k-1)\odot_{G}m=n$, and
\beqa \ket{\Psi}^{k}_{(i_{1},i_{2})}={1\over N^2}(\sum_{m=0}^{N-1}
\gamma_{G}^{(i_{1},i_{2})\odot\odot_{G}(m,(k-1)\odot_{G}m)}\ket{B^k_{0^*,m}})\nonumber \\
 +{\rm contributions\ of\ weigth\ 0}\nonumber \\
  k=1...N
\nonumber\eeqa
The last expression is similar to the one obtained in the computational basis. This proves the ``invariance'' of the relevant components
 of the states $\ket{\Psi}$. By a computation similar to the one performed
  in the computational basis, we get that 
 $\braket{\Psi^{k}_{(i_{1},i_{2})}}{  e^k_{l^*Alice}  e^k_{lKing}}=
   {1\over \sqrt N}\delta_{ i_{1}\oplus_{G} (i_{2}\odot_{G} (k-1)\odot_{G} R),
   \ominus_{G} l}$. 
   Once again, this relation allows Alice to infer the label $l$ of the King's observation/preparation 
   unambiguously from the
    labels $ i_{1}, i_{2}$ of the detector that fires.
    
    We can understand better the invariance of the states $\ket{\Psi}$ if we 
    note that when we pass from the computational basis to any of the corresponding MUB's,
     the Bell states transform in such a way that
     
      $(\gamma_{G}^{  (  m\odot_{G} n )})^{1\over 2}
     \ket{B^0_{(m^*,n}}=(\gamma_{G}^{(\ominus_{G} m\odot_{G} n)})^{1\over 2} 
 .(\gamma_{G}^{  ((k-1)\odot_{G}  m\odot_{G} m )})^{1\over 2}
\ket{B^k_{(\ominus_{G}n\oplus_{G}(k-1)\odot_{G}m)^*,m}}$
     
     In odd prime power dimensions, $¥(\gamma_{G}^{(\ominus_{G} m\odot_{G} n)})^{1\over 2} 
 .(\gamma_{G}^{  ((k-1)\odot_{G}  m\odot_{G} m )})^{1\over 2}
\ket{B^k_{(\ominus_{G}n\oplus_{G}(k-1)\odot_{G}m)^*,m}}$¥$¥=
(\gamma_{G}^{  (  m'\odot_{G} n' )})^{1\over 2}
     \ket{B^k_{(m'^*,n'}}),$ with 
     
     \begin{equation}
\label{solschrod} \left( \normalbaselineskip=36truept\matrix{
m'\cr
n'\cr
}\right)=\left( \normalbaselineskip=36truept
\matrix{
 (k-1)&\ominus_{G}1\cr
1&0\cr
}\right)\left( \normalbaselineskip=36truept\matrix{
m\cr
n\cr
}\right)
 \end{equation}¥
     
  This transformation is remarkable in the sense that it preserves 
  the symplectic form $m_{1}\odot_{G}n_{2}\ominus_{G}n_{1}\odot_{G}m_{2}$. Indeed, as
   a consequence of the transformation law (\ref{solschrod})¥, this form is the same in all MUB's:¥
  $m'_{1}\odot_{G}n'_{2}\ominus_{G}n'_{1}\odot_{G}m'_{2}=
  m_{1}\odot_{G}n_{2}\ominus_{G}n_{1}\odot_{G}m_{2}¥¥¥$¥.
  
  In odd prime dimensions, where the Galois operations reduce to the modulo $N$¥ operations, this
   property was intensively studied,\cite{Appleby,davidsThesis,Vourdas} as well as its numerous applications. Here we see that 
  these properties also generalise in odd prime power dimensions. 
  
  It is possible to explicit the invariance of the states $\ket{\Psi}^{k}_{(i_{1},i_{2})}$ more elegantly, by expressing it directly in terms of the 
  conserved symplectic form. This can be done by relabelling them as follows:
  
  $¥¥\ket{\Psi'}^{0}_{(i_{1},i_{2})}={1\over N}(\sum_{m,n=0}^{N-1}
  \gamma_{G}^{(i_{2}\odot_{G}m\ominus_{G}i_{1}\odot_{G}n)} (\gamma_{G}^{( m\odot_{G} n)})^{1\over 2} 
  \ket{B^{0}_{m,n}}_{A,K})$¥
  
  This relabelling is bijective because  
  
  $\ket{\Psi'}^{0}_{(i_{1},i_{2})}=
  \ket{\Psi}^{0}_{(i_{2},\ominus_{G}i_{1}/_{G} R)}$¥¥ and $R$,
   the remainder of $(0,1)\odot\odot_{G}(0,1)$ 
   after division by $N$ always differs from 0, as a consequence of the fact that a quadratic extension of a field is a field. 
 
 The transformation law for the relabelled states is 
 
 \begin{equation} \ket{\Psi'}^{k}_{(i'_{1},i'_{2})}=\ket{\Psi'}^{0}_{(i_{1},i_{2})}\label{essential}\eeq
 where the transformation law between primed and unprimed paris of indices obeys the bijective transformation law (\ref{solschrod}). 
 
 Unfortunately this elegant transformation law is not respected in general in even prime dimensions, as shows our example in the qubit case where one can check for oneself that the 
 four $\Psi$ states of Alice's basis do not exactly transform into each other when they are reexpressed into different MUB's. This is also the case for the $U$'s operators (the Pauli operators in this case) which are 
 permuted among each other up to multiples of minus 1 (the additive character of the field). We mentioned already that such
  phases can be compensated
  by a relabelling of the basis states. Indeed, the group structure is preserved under conjugation so that
   the phases of the $U$ operators expressed in a different basis may be compensated by a Galois shift of the label of the basis states.
  Nevertheless this is true for subgroups of the full Pauli group only and it is impossible to compensate all 
 phases. In other words no relabelling of the basis states is consistent with the requirement that
  $¥(\gamma_{G}^{(\ominus_{G} m\odot_{G} n)})^{1\over 2} 
 .(\gamma_{G}^{  ((k-1)\odot_{G}  m\odot_{G} m )})^{1\over 2}
$=¥$¥(\gamma_{G}^{  (  m'\odot_{G} n' )})^{1\over 2}$ with $(m',n')¥$ defined by Eq.~(\ref{solschrod}¥)) for all couples $(m,n)¥$ SIMULTANEOUSLY, by a single relabelling of the basis states.
 For instance if we reorder the spin up and down states along the $X$ direction 
 (which means that we perform a spin flip) we change the sign of two Bell states (in the second terms of the inequalities (\ref{bla}¥))¥ and not only one as required. 
This is a simple example which shows that the symplectic structure is invariant in odd prime power dimensions only
\footnote{Recently, D. Gross drew our attention on the fact that this property is also true in arbitrary dimensions,
 when we work with the modulo $N$ operations: the even dimensional case does not present the same guarantee
  regarding the symplectic invariance as the odd one.
¥}.¥¥¥¥¥¥¥¥

 \section{Connection with the discrete Weyl and Wigner distributions.}
 
\noindent
 We showed that, up to a phase, generalised Bell states are in one to one correspondence with generalised displacement operators
  (the $V$ or $U$ operators). The $N^2$ displacement opertors form a basis of the space of linear $N$x$N$ operators, and any density matrix can be 
 expressed as a linear combination of the displacement operators, which presents interesting consequences concerning tomographic
  applications. As the displacement operators are diagonal in the MUB's, there is a direct connection between MUB's and tomography, a fact that was already recognised and studied in depth in the past.\cite{Wootters,Ivanovic} 
  
  Here we shall focus on the interrelations between our results and discrete quasi-distributions. In odd prime dimensions, 
  the Galois and modulo $N$ operations coincide, and the discrete version of Weyl's and Wigner's distributions
   is well-known in that case.\cite{Vourdas} The identification with our results is straightforward. For instance, the displacement operators
    defined in Eqn. (28) of Ref.\cite{Vourdas2} coincide with the $U$ operators defined in our paper, when the dimension is odd and prime.  
   The Weyl function is intimately
    related to the displacement operators and can thus be expressed in terms of the $U$ operators. The Weyl function of a linear $N$x$N$ operator $O$ ¥is defined in Ref.\cite{Vourdas2} (Eqn.82) by the relation

     $\tilde{W}(O,\alpha,\beta)=
    \gamma^{{-_{mod.N}\alpha._{mod.N}\beta/_{mod.N} 2}}\sum_{l=0}^{N-1}\gamma^{-_{mod.N}\beta ._{mod.N} l}
    \bra{  e_{l}^0}O \ket{  e_{\alpha+_{mod.N}l}^0}$ (valid in odd prime dimensions). It is easy to check that $Tr.U^i_{l}.O=
    \tilde{W}(O,l,(i-1)._{mod.N}l)$, so that the Weyl's function corresponding to the operator $O$ is nothing else than the amplitude
     of this operator when it is expanded in the $U$ operators basis. This is also true when $O$ is a density matrix and can be
      generalised in a straightforward manner to prime power dimensions (replacing modulo $N$ operations by Galois operations¥)¥. In this perspective, 
      our non-standard generalisation of the displacement operators to prime power dimensions 
    provides a discrete version of  ``\`a la Galois'' Weyl's quasi-distribution. 
    
    As we noted before there is a one to one correspondence between Bell states and displacement operators.
     This correspondence works in both directions and we can thus associate to the basis $\ket{\Psi'}^{0}_{(i_{1},i_{2})}$ 
     of the qu$N^2$it space a basis of linear $N$x$N$ operators. As we shall now show, the amplitudes of the expansion of a linear$N$x$N$ operator $O$
      in this basis are, in odd prime dimensions, equivalent to the Wigner function $W(O,\alpha,\beta)$ defined in Ref.\cite{Vourdas2}. Expressing by the symbol $\Psi^{0}_{(i_{1},i_{2})}$ 
   the operator obtained from $\ket{\Psi'}^{0}_{(i_{1},i_{2})}$ by replacing formally Alice's conjugate kets by bras, we get that 
   $W(O,\alpha,\beta)=Tr.\Psi^{0}_{(i_{1},i_{2})}.O=(1/N)¥\sum_{\alpha,\beta=0}^{N-1}
   \gamma^{\alpha ._{mod.N} i_{2}
   -_{mod.N}\beta ._{mod.N} i_{1}}\tilde{W}(O,\alpha,\beta)$ in agreement with the relation (83)¥ in Ref.\cite{Vourdas2}. 
   This is also true when $O$ is a density matrix so that 
     our non-standard generalisation of the  $\Psi^{0}_{(i_{1},i_{2})}$ operators to prime power dimensions 
    provides a discrete version of Wigner's quasi-distribution. 
    
    In odd prime dimensions, it is also possible to express Wigner operators in terms of the parity operator $P_{0,0}$ and of 
    the displacement operators, in accordance with the expressions (28), ¥(61) and (66) of Ref.\cite{Vourdas2} that can be respectively 
    rewritten, according to our conventions, as follows:
    
     $D¥(\alpha=l,\beta=(i-1)\odot_{mod. N}l)
    =(\gamma_{mod. N}^{\ominus_{mod. N}((i-1)\odot_{mod. N} l\odot_{mod. N} l)})^{1\over 2}V^{(i-1)\odot_{mod. N}l }_{l}
    =U^{i}_{l}  (V28 ¥)¥$,
    
     $P(\alpha,\beta)=
    D(2\odot_{mod. N}\alpha,2\odot_{mod. N}\beta)¥P(0,0)¥ (V61)¥$ and 
    
    $P(0,0)=¥{1\over N}\sum_{¥\alpha,\beta=0}^{N-1}
    D¥(\alpha,\beta) (V66)¥$¥¥ where the labels V in the numbering of the previous expressions refers to corresponding equations in Prof. Vourdas's paper.\cite{Vourdas2}¥
    
    The Mean King operators $\Psi^{0}_{(\alpha,\beta)}$ can be written under the form 
    $\Psi^{0}_{(\alpha,\beta)}=\sum_{¥\tilde \alpha,\tilde \beta=0}^{N-1}$¥
    $(\gamma_{mod. N}^{\ominus_{mod. N}\alpha\odot_{mod. N} \tilde\beta\oplus_{mod. N}
    \tilde\alpha\odot_{mod. N} \beta})$
    $(\gamma_{mod. N}^{\ominus_{mod. N}(\tilde\alpha\odot_{mod. N} \tilde\beta)})^{1\over 2}$
    $¥¥¥    V¥^{\tilde \beta}_{\tilde \alpha}$¥¥. Reexpressing the variables $(\tilde \alpha,\tilde \beta)$¥ via the substitution 
    $¥\tilde \alpha=2 \odot_{mod. N} \alpha\oplus_{mod. N}\alpha'$ and $¥\tilde \beta=2  \odot_{mod. N}\beta\oplus_{mod. N}\beta'$¥ 
    we get after some simplifications the relation 
    
    $\Psi^{0}_{(\alpha,\beta)}=
    \sum_{¥ \alpha', \beta'=0}^{N-1}(\gamma_{mod. N}^{\ominus_{mod. N}(2\alpha\odot_{mod. N} 2\beta)})^{1\over 2}¥$¥
    $(\gamma_{mod. N}^{\ominus_{mod. N}(\alpha'\odot_{mod. N} \beta')})^{1\over 2}$¥
    $(\gamma_{mod. N}^{\ominus_{mod. N}(2\alpha\odot_{mod. N} \beta')})V^{\tilde\beta}_{\tilde \alpha}$, with $¥\tilde \alpha=2 \odot_{mod. N} \alpha\oplus_{mod. N}\alpha'$ and $¥\tilde \beta=2  \odot_{mod. N}\beta\oplus_{mod. N}\beta'.$¥

     The righthand term of the previous equality is in turn equivalent to  the displaced parity operator defined through the relation 
    (61)  in Ref.\cite{Vourdas2} ¥(V61) so that the Mean King operators are equal to the displaced parity operators of Refs.\cite{Vourdas,Vourdas2}: $\Psi^{0}_{(\alpha,\beta)}=P_{(\alpha,\beta)}$, when the dimension is odd and prime¥.¥¥
    
It is worth noting that, although the expression of the Mean King operators in terms of the displaced parity operator can be generalised in a straightforward manner to the odd prime power dimensional case, 
this is no longer true in even prime power dimensions because the parity operator is then equal to the identity, so that the $P$ operators defined through the 
relation $(V61)¥$¥ coincide with the displacement or Weyl operators¥ \footnote{Prof. Paz signalled this problem
 to me during a conversation that we had at the ICSSUR conference hold
 in Besancon in May 2005}. Nevertheless, 
the Mean King operators or generalised Wigner operators can still be defined, also in even prime power dimensions through the relation

\beqa \Psi^{0}_{(i_{1},i_{2})}={1\over N}\sum_{m,n=0}^{N-1}
  \gamma_{G}^{\ominus_{G}i_{1}\odot_{G}n \oplus_{G}i_{2}\odot_{G}m} (\gamma_{G}^{( m\odot_{G} n)})^{1\over 2} V^{0}_{m,n}
  \eeqa
  In even prime power dimensions, we obtain via this relation good candidates for a discrete Wigner distribution that differ from the discrete Weyl distribution. For instance in dimension 2 we obtain 
  that the four Wigner operators are equal to the 4x4 Hadamard transform of the Pauli operators:
  
   $2\Psi^{0}_{(0,0)}=(¥¥\sigma_{0,0}+¥¥\sigma_{1,0}+\sigma_{0,1}+\sigma_{1,1})$¥,
   
   $2\Psi^{0}_{(0,1)}=(¥¥\sigma_{0,0}-¥\sigma_{1,0}+\sigma_{0,1}-\sigma_{1,1})$¥,
   
   $2\Psi^{0}_{(1,0)}=(¥¥\sigma_{0,0}+¥¥\sigma_{1,0}-\sigma_{0,1}-\sigma_{1,1})$¥,
   
   $2\Psi^{0}_{(1,1)}=(¥¥\sigma_{0,0}-¥¥\sigma_{1,0}-\sigma_{0,1}+\sigma_{1,1})$¥,

  (with  $\sigma_{0,0}=Identity=\ket{0}\bra{0}+\ket{1}\bra{1} $,  $\sigma_{0,1}=
  \sigma_{Z}¥ =\ket{0}\bra{0}-\ket{1}\bra{1}$, $\sigma_{1,0}=\sigma_{X}¥ =
  \ket{0}\bra{1}+\ket{1}\bra{0}$¥¥¥ and $\sigma_{1,1}=\sigma_{Y}¥=-i\ket{0}\bra{1}+i\ket{1}\bra{0} $.¥
       
   One can check that the marginals exhibit the behavior that we are in right to expect from well-defined
    Wigner distributions; for instance, 
    $\Psi^{0}_{(0,0)}+\Psi^{0}_{(0,1)}=2\ket{0}\bra{0}$¥ 
    and $\Psi^{0}_{(1,0)}+\Psi^{0}_{(1,1)}=2\ket{1}\bra{1}$. This qubit discrete
     Wigner function corresponds to the Wootters qubit phase-space distribution\cite{Wootters2} and is directly
      related to the tetrahedron qubit technique of tomography developed
       at NUS by Englert and coworkers.\cite{tetra,tetra'} In prime power dimensions, the generalisation is straightforward because 
    
    $N\sum_{\beta=0}^{N-1}\Psi^{0}_{(\alpha,\beta)}=
   \sum_{\beta=0}^{N-1} \sum_{¥\tilde \alpha,\tilde \beta=0}^{N-1}$¥
    $(\gamma_{G}^{\ominus_{G}(\alpha\odot_{G} \tilde\beta\oplus_{G}\tilde\alpha\odot_{G} \beta)})$
    $(\gamma_{G}^{\ominus_{G}(\tilde\alpha\odot_{G} \tilde\beta)})^{1\over 2}$
    $¥¥¥    V¥^{\tilde \beta}_{\tilde \alpha}$¥

    $=
    \sum_{¥\tilde \alpha,\tilde \beta=0}^{N-1}$¥
    $N.\delta_{\tilde\alpha,0}(\gamma_{G}^{\ominus_{G}
    (\alpha\odot_{G} \tilde\beta)})$
    $(\gamma_{G}^{\ominus_{G}(\tilde\alpha\odot_{G} \tilde\beta)})^{1\over 2}$
    $¥¥¥    V¥^{\tilde \beta}_{\tilde \alpha}$¥
    
    $=
    \sum_{¥\tilde \alpha,\tilde \beta=0}^{N-1}$¥
    $N.(\gamma_{G}^{\ominus_{G}
    (\alpha\odot_{G} \tilde\beta)})$ $¥¥¥    V¥^{\tilde \beta}_{0}$¥
    
    $=
    N^2.\ket{  e^{0}_{\alpha}}\bra{  e^{0}_{\alpha}} .$¥

    ¥We can obtain similar results in the $X$ and $Y$ bases by realising the two other possible splittings of the set of four 
    Wigner operators into two pairs of operators. These splittings correspond to the concept of striation 
    that was studied in depth in Refs.\cite{Wootters2,discretewigner,paz}.
   Each striation can be put in one to one correspondence with a group of $N-1$ commuting displacement operators plus the identity or, equivalently, with the associated MUB in which this group is diagonal. 
   The marginal property of the discrete Wigner distribution can thus be rewritten
    in any of the $N+1$ MUBs, not only in the computational and dual bases. This is easily established
     making use of the identities 
   (¥\ref{transfowigner}¥¥)¥ or (\ref{essential})¥. 
   
   It is worth noting that in our approach the ambiguities related to the derivation of a discrete phase space are implicitly evacuated from the beginning because we impose an a priori order to the MUBs (and thus to the associated striations). This particular choice is a consequence of our choice of the basis for the 
 Galois field, of our identification between each element of the field and a state 
 of the computational basis, of our particular phase choice for the $U$ operators and of our numbering of the $N+1$ families of commuting displacement operators. 
 All our choices were fully arbitrary, which reflects 
 our intimate conviction that there is no prefered manner to order the set
  of MUB's or to order the states inside a given basis (up to a Galois translation). 
  We systematically chose the ordering that seemed to be the most convenient and natural. 
  In last resort, such a choice ought to be dictated by external arguments and symmetries;
   for instance if we identify the $N$ dimensions with a physical parameter like a discretised position or 
   angle, then the physical realisation of the states imposes a natural ordering, which was not the case
    in our approach.
    
    In any case, the transformation law (\ref{essential}) shows that, in odd prime power dimensions, an essential
     invariance is guaranteed: the phase space picture is the same (up to a bijective relabelling)
      whenever¥¥ we pass from the computationals basis to any of the $N$¥mutually unbiased bases
       defined in Eq.~(\ref{synthetic})¥. In even prime power dimensions,
        the symmetry is less strong because phases plus and minus appear that cannot be eliminated as we noted before, but our solution of the Mean King's problem shows that anyhow
         some regularity is still present in this case. It is worth noting another difference between the odd and even dimensional Wigner distributions presented here: the displaced parity operators (which are the 
         odd dimensional Wigner operators)¥ are always factorisable in all MUB's into product of local operators. This is not necessarily true for the Wigner operators in even prime power ($2^m$¥) ¥dimensions.

 \section{Connection with Aravind's general solution.} 
 
\noindent
 P.K. Aravind has shown\cite{2003} how to generalise to prime power dimensions the solution of Aharonov and Englert, valid in
  prime dimensions, to prime power dimensions. It is not our goal to compare in detail our 
  approach and his approach but rather to situate the approaches relatively to each other. 
  The expression of Aravind for the Mean King's basis is very general because it can be expressed in terms of the
   $N+1$ ¥mutually unbiased bases,
   independently on how they were derived. It is also essentially unique, up to a permutation of the state labels or a rearrangement in their relative positions. 
 Our solution is less general because we chose a particular expression and ordering for the MUB's;
  nevertheless, it must coincide with Aravind's expression 
 once this choice is made because Aravind's states can be shown to be orthogonal 
 to the states $\ket{  e^{m*}_{k}e^{m}_{k}}_{A,K}¥$ with $k(m)\not=(m-1)k_{0}+k_{1}¥¥¥¥$,
  when $m\not=0$ and $m:1...N$ and $k(0)\not=k_{0}¥¥¥¥$,
  when $m=0$.¥¥ This constraint defines unambiguously the 
 state that is labelled by a pair of indices $k_{0},k_{1}$¥ (up to a global phase). One can check that the states 
 $¥¥\ket{\Psi}^{0}_{(i_{1},i_{2})}$¥ defined in Eq.~(\ref{alicestates}) and correspond to our solution of the Mean King's problem fulfill exactly the same constraints, which establishes the 
 convergence of both approaches. Nevertheless, it is worth noting that the additive character that appears in 
 Aravind's expressions¥¥ is not the additive character of the Galois field that we 
 considered ($p$th root of unity¥)¥¥ but the additive character of the modulo $N$ ring ($N=p^m$th root of unity¥).¥  Anyhow, unicity of the solution to Aravind's constraints imposes that both expressions must coincide.

\section{Conclusion}

\noindent

We showed in another paper\cite{Durtmutu} how the construction of the MUB's can be derived from very primitive concepts: addition, multiplication, and duality. The whole structure can be derived in a self-consistent manner, making use of some well-known properties exhibited by finite fields. 
The basic intuition  that finally led to the resolution of the Mean King's problem is the recognition that the generalised Pauli group (sometimes called Heisenberg-Weyl group) is nothing else than a discrete Fourier transform performed on the elements of a Galois field. 
All the rest can be derived on the basis of this simple property. The result shows the power of entanglement seen as a
 resource: certain tasks that would be impossible if we remained confined to a $N$-dimensional Hilbert space because they 
 would contradict the uncertainty principle are possible provided we add an ancilla and exploit the counterintuitive properties offered by entanglement.

Although our solution can be considered merely as a special case of Aravind's general solution,\cite{2003}
  it exhibits interesting properties, due to our particular way of expressing mutually unbiased bases. It provides
   among others an interesting generalisation of the discrete Wigner function
   that was proposed in prime dimensions in the past.\cite{Vourdas,Vourdas2} It is an open question to know whether generalised Wigner functions could be associated to arbitrary set of MUB's in the same sense that Aravind's solution for the Mean King's problem generalises our solution. Finally our approach allows us to generalise the Mean King's problem in odd non-necessarily prime dimensions as we show in appendix.

\section*{Acknowledgements}

\noindent

The author acknowledges
a Postdoctoral Fellowship of the Fonds voor Wetenschappelijke Onderzoek,
Vlaanderen and also support from the
IUAP programme of the Belgian government, and the grant V-18.  
thank you to Prof.J. Corbett for his kind hospitality at the Macquarie university of Sydney where a part of this 
work has been completed. Many thanks to Profs. Vourdas, Paz, Klimov, Planat, Appleby, Gross and De Guise for discussions and comments
 and also to
 B-G Englert (NUS) for fruitful explanations about the Mean King's problem.

\appendix

{\bf Reformulation of the Mean King's problem in odd dimensions.}

Although the modulo $N$ operations form a field only when the dimension is prime, many properties that were valid 
when we made use of the Galois operations can be generalised in terms of the modulo $N$ operations. Because of this, we can 
generalise the Mean King's problem in arbitrary odd dimensions as follows: Alice must be able to ascertain 
    the value of $N+1$ (non-necessarily) commuting (and possibly degenerated) observables. 
    The observables are the displacement operators previously introduced but now they must be reexpressed in terms of the $N$th root of unity (instead of $p$¥) and of the modulo $N$ operations.¥¥¥ They can be shown by a derivation entirely similar to the chain of equalities Eq.~(\ref{zzz}) to
     be
     diagonal in the 
    $N+1$ (non  necessarily mutually unbiased) bases defined as follows:
  \begin{equation} 
  \ket{  e_{k}^i}={ 1\over \sqrt N}\sum_{q=0}^{  N-1}\gamma_{N}^
  { \ominus_{mod. N} q\odot_{mod. N} k}
  (\gamma_{N}^{(
  (i-1)\odot_{mod. N} q\odot_{mod. N} q)/_{mod. N}2}) \ket{  e_{q}^0}
\label{that} \end{equation}   
  where $\gamma_{N}$ represents a $N$th root of unity. As usually the 0th basis
   is the computational basis. The solution of the generalised Mean King's problem in arbitrary odd dimensions ¥is entirely similar to the case where dimensions are prime powers, excepted that 
   we must replace the Galois operations between $N$ elements by the modulo $N$ operations and the 
   product $(i_{1},i_{2})\odot\odot_{G}(m,n)$ by 
   $\ominus_{mod. N}(i_{1}\odot_{mod. N}n)\oplus_{mod. N}(i_{2}\odot_{mod. N}m)$. All the transformation 
   properties of the Bell states when we pass from one MUB to another one are preserved and the full reasoning 
   can be repeated integrally. The main difference is that the $N+1$ 
   ¥eigenbases described by the expression (\ref{that})
    are no longer mutually unbiased 
    (we conjecture that at most $p+1$ of them are mutually unbiased, where $p$ is the smallest prime divider
     of $N$).


\section*{References}
\begin{thebibliography}{99}
\bibitem{DurtNagler} B. Nagler and T. Durt, Phys. Rev. A {\bf 66} 042323 (2003).
\bibitem{Durtmutu} T. Durt, quant-ph/0401046, 1-24 (2004).
\bibitem{Wootters} W.K. Wootters, and B.D. Fields, Ann. Phys. {\bf 91}, 363 (1989).
\bibitem{india}S. Bandyopadhyay, P. Boykin, V. Roychowdhury, and F. Vatan, Algorithmica {\bf 34}, 512 (2002) 
(quant-ph/0103162, 1-22 (2001)).
\bibitem{Schwinger}J. Schwinger, Proc. Nat. Acad.Sci. U.S.A., {\bf 46} 570 (1960).
\bibitem{Ivanovic} I.D. Ivanovic, J. Phys. A, {\bf 14}, 3241 (1981).
\bibitem{Durtsept} T. Durt, J. Phys. A: Math. Gen. 38 (2005) 5267-5283.¥
\bibitem{vaid} L. Vaidman, Y. Aharonov, and D.Z. Albert, Phys. Rev. Lett. {\bf 58}  1385 (1987). 
\bibitem{Englert} Y. Aharonov and B.-G. Englert, Z. Naturforsch. {\bf 56}, 16 (2001). 
\bibitem{2003} P.K. Aravind, Z. Naturforsch. {\bf 58A} 2212 (2003).
\bibitem{klimov} A. B Klimov, L. L Sanchez-Soto and H. de Guise, J. Phys. A. {\bf 38} 2747 (2005).¥
 \bibitem{Karpilovsky}G. Karpilovski, {\it Field theory} (Marcel Dekker Inc. New-York-Basel, 1988).
\bibitem{weyl}H. Weyl, Z. Phys. {\bf 46} 1 (1927), H. Weyl: {\it Gruppenthorie und Quantemechanik } (1928), english translation by H.P. Robertson, E.P. Dutton, N-Y (1932).
 \bibitem{Planat}M. Planat, H. Rosu, S. Perrine and M. Saniga, quant-ph/0409081, 1-14 (2004).
\bibitem{DurtKwek}T. Durt, D. Kaslikowski, J-L Chen, en L.C. Kwek, quant-ph 0302078, 1-15 (2003).
 \bibitem{Appleby} M. Appleby, quant-ph/0412001, 1-26 (2004).
 \bibitem{davidsThesis}
D. Gross, {\it Phase space methods in quantum information.} Diploma
Thesis, University of Potsdam 2005. Available online at
http://gross.qipc.org.
  \bibitem{DurtMean} T. Durt, quant-ph/0401037, 1-10 (2004).
 \bibitem{Rubin}A.O. Pittenger and M.H. Rubin, quant-ph/0308142, (2003).
 \bibitem{Vourdas} A. Vourdas, J. Phys. A: Math.Gen. {\bf 29}, 4275 (1996).
 \bibitem{Vourdas2} A. Vourdas, Rep. Progr. Phys. {\bf 67}, 267 (2004).
 \bibitem{Wootters2} W-K Wootters, Ann. of Phys.¥ {\bf 176}, 1 (1987), W.K. Wootters, quant-ph/0406032,¥ 1-26  (2004).
\bibitem{tetra}J. Rehacek, B-G Englert, D. Kaszlikowski, Phys. Rev. A {\bf 70}, 052321 (2004).
\bibitem{tetra'}A.E. Allahverdyan, R. Balian and Th. Nieuwenhuizen, Phys. Rev. Lett. {\bf 92}, 12, 120402.
\bibitem{discretewigner} K.S. Gibbons, M.J. Hoffman and W.K. Wootters,
 quant-ph/0401155, 1-60 (2004).
 \bibitem{paz}J. Paz, A. Roncaglia, and M. Saraceno, quant-ph/0400117, 1-19 (2004).


\end{thebibliography}
\end{document}